\documentclass[conference]{IEEEtran}\IEEEoverridecommandlockouts
\usepackage[left=1.62cm,right=1.62cm,top=1.9cm]{geometry}
\usepackage{cite}
\usepackage{amsmath}
\usepackage{amssymb}
\usepackage{graphicx}
\usepackage{subfigure}
\usepackage{caption}

\begin{document}

\title{
Enhancing Vital Sign Estimation Performance of FMCW MIMO Radar by Prior Human Shape Recognition
}
	\author{
		\IEEEauthorblockN{Hadi Alidoustaghdam$^{*1}$, Min Chen$^\dagger$, Ben Willetts$^*$, Kai Mao$^*$, André Kokkeler$^*$, Yang Miao$^*$}\\
		\IEEEauthorblockA{$^*$Faculty of Electrical Engineering, University of Twente, Enschede, The Netherlands\\
  $^\dagger$School of Electronics and Information Engineering, Harbin Institute of Technology, Harbin, China}\\
			\emph{Email$^1$}: hadi.alidoustaghdam@utwente.nl}

\maketitle

\begin{abstract}
 Radio technology enabled contact-free human posture and vital sign estimation is promising for health monitoring. Radio systems at millimeter-wave (mmWave) frequencies advantageously bring large bandwidth, multi-antenna array and beam steering capability. \textit{However}, the human point cloud obtained by mmWave radar and utilized for posture estimation is likely to be sparse and incomplete. Additionally, human's random body movements deteriorate the estimation of breathing and heart rates, therefore the information of the chest location and a narrow radar beam toward the chest are demanded for more accurate vital sign estimation.

In this paper, we propose a pipeline aiming to enhance the vital sign estimation performance of mmWave FMCW MIMO radar. 
The first step is to recognize human body part and posture, where we exploit a trained Convolutional Neural Networks (CNN) to efficiently process the imperfect human form point cloud.
The CNN framework outputs the key point of different body parts, and was trained by using RGB image reference and Augmentative Ellipse Fitting Algorithm (AEFA). 
The next step is to utilize the chest information of the prior estimated human posture for vital sign estimation. While CNN is initially trained based on the frame-by-frame point clouds of human for posture estimation, the vital signs are extracted through beamforming toward the human chest. 

The numerical results show that this spatial filtering improves the estimation of the vital signs in regard to lowering the level of side harmonics and detecting the harmonics of vital signs efficiently, i.e., peak-to-average power ratio in the harmonics of vital signal is improved up to 0.02 and 0.07dB for the studied cases. 
\end{abstract}


\section{Introduction}
Human health monitoring in home and work environments is critical and demands reliable, quick, contactless and privacy preserved methods \cite{le2019radar}. In an outlook on 6G and ubiquitous wireless devices, radio based sensing, with high bandwidth and numerous antennas, is a reasonable technology \cite{wei2022toward,9330512}. However, its reliability is still under doubt, i.e., the point clouds generated by millimeter-wave (mmWave) radars are sparse and often not informative of human's posture per se \cite{sengupta2020mm}, or removing the effects of random body motions (RBMs) for vital sign estimation, i.e., breathing rate (BR) and heart rate (HR), requires spatial filtering and signal processing stages \cite{dai2021enhancement,singh2020multi,gouveia2019review}. When a human is considered as an extended target in the field of view of a radar, or multiple human monitoring cases \cite{xu2022simultaneous},  proper range and cross-range bin selection are crucial for vital sign extraction. The majority of commercial radars usually have a limited number of antennas and then limited cross-range resolution, which complicates human sensing.
Although, signal processing and machine learning techniques are applied to alleviate this deficiency in cross-range resolution \cite{singh2020multi}, there is still potential in the measurement process through distributed radars \cite{iwata2021multiradar}, beam steering,  and sensor fusion \cite{shokouhmand2022camera}.
The following two papers are very relevant to our work in this paper. First, A. Sengupta \textit{et al}\cite{sengupta2020mm} proposed to estimate skeleton joints by projecting
the radar reflection points onto the depth-azimuth and depth-elevation planes, respectively, and then train a convolutional neural network (CNN) architecture to estimate the 3D coordinates of skeleton joints. 
Second, the authors in \cite{shokouhmand2022camera} suggest that the beam steering of measurements of the mmwave radar toward the chest can be applied to extract vital signs. The ground-truths for both papers are provided by Kinect which is a combination of RGB camera and infra-red (IR) depth sensor.

 Vibrations of a human chest wall due to breathing and heartbeat can be detected via modulated  reflected signals of frequency-modulated continuous-wave (FMCW) radar. Among vital signs, e.g., body temperature, blood pressure, breathing, and heart rate, the goal of this paper is to enhance the BR and HR estimations with prior information on posture and body parts; hence, a pipeline of human posture and vital sign estimation using CNN and mmWave multiple-input-multiple-output (MIMO) radar is proposed. The dependence on camera images (as ground truth references) is only limited to the training phase of the CNN. MmWave radar measurements contain depth information, but with lower cross-range resolution than IR-depth sensor; however when the radar has a full front view of the human, which is the assumption of this paper, a CNN can compensate for this deficiency in cross-range resolution. The contributions of this paper are twofold. 
\begin{itemize}
    \item First, to advance contact-free human posture recognition, this paper proposes to leverage a CNN to estimate human postures and model human shapes. Note that the shape contains the information about the size of various body parts. The Augmentative Ellipse Fitting Algorithm (AEFA) \cite{panagiotakis2016parameter} is applied on the RGB image to estimate a human's shape by ellipses and pass this information as ground truth to train a CNN. The trained CNN then estimates the human's posture based on the imperfect point cloud captured by a mmWave MIMO FWCW radar.
    \item Second, we show how the posture estimation is beneficial for vital sign extraction of a human. It is shown that beamforming toward chest, which is already estimated by  the trained CNN, can increase the signal to noise and interference ratio and finally improve the performance of vital sign estimation. 
\end{itemize}

\begin{figure}[h]
	\centering{\includegraphics[width=0.8\linewidth]{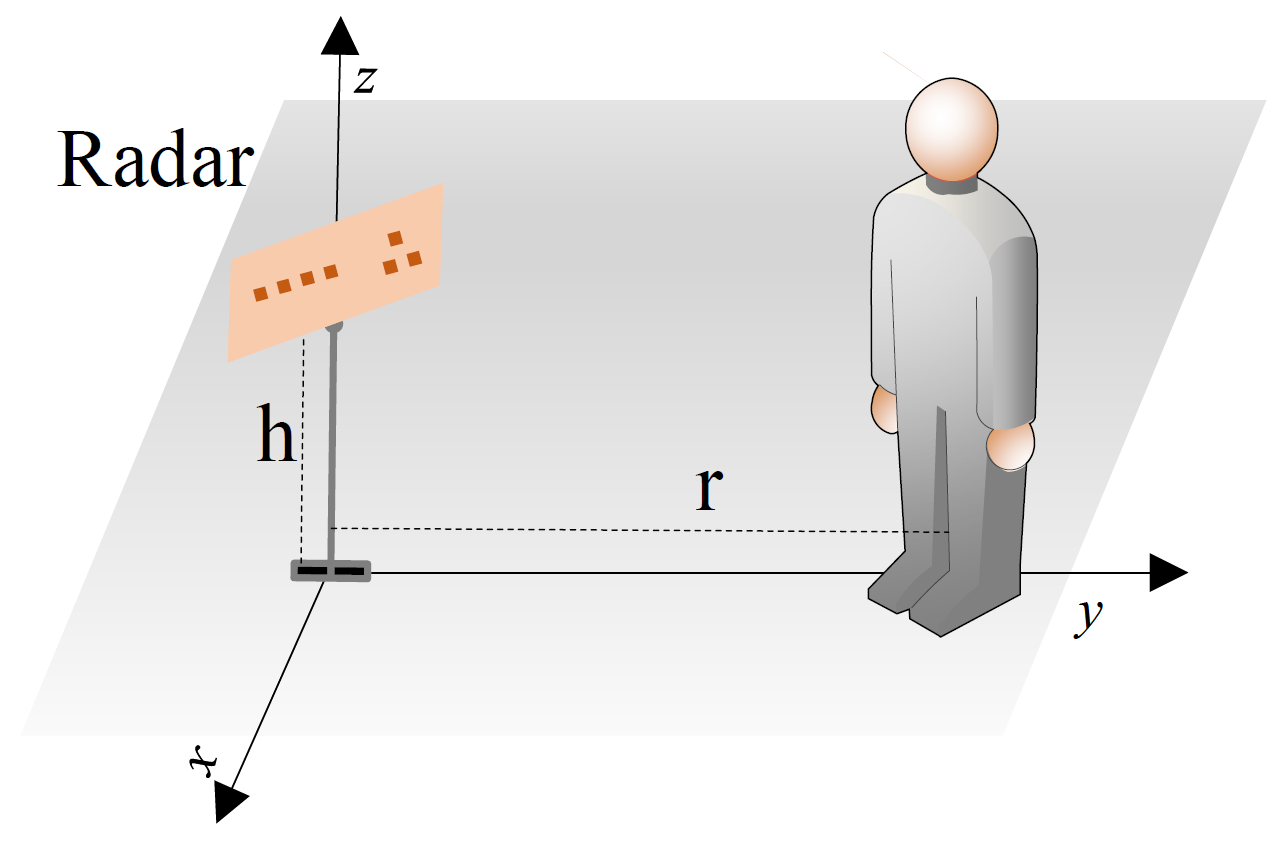}}
	\caption{A human standing in front of FMCW radar.}
	\label{scenario}
\end{figure}
\section{The Pipeline for Human's Posture  and Vital Sign Estimation}
As there are appropriate explanations of the working principles of FMCW MIMO radar in, e.g., \cite{abedi2021ai,alizadeh2019remote}, we will skip it and only provide the parameters of the measurement. However, it is necessary to mention that the output of the FMCW MIMO radar is a 3D matrix which, after signal processing,  contains information on the range, velocity, and cross-range of targets. The resultant output is usually a sparse 3D point cloud frame by frame. Fig.~\ref{scenario} sketches the measurement scenario. 

\subsection{Human Body Parts and Posture Identification}
\begin{figure*}
\centering
\includegraphics[width=0.9\linewidth]{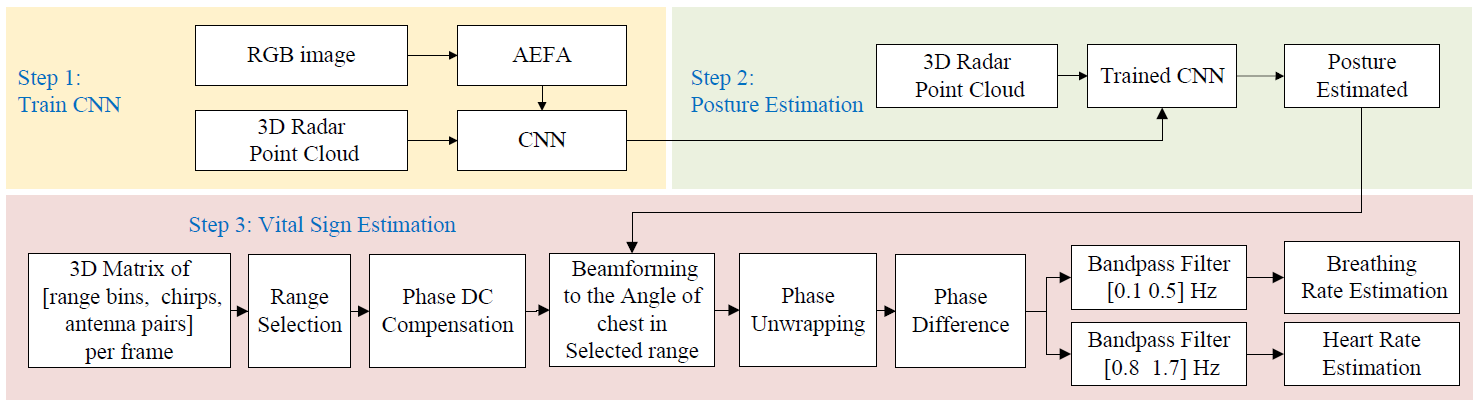}
\caption{The proposed pipeline for consecutive human posture/shape and vital sign estimation.}
\label{PipeLine}
\end{figure*}
Here we elaborate on how several key points/features of different body parts can be estimated and then stitched together to recognize human postures by using the AEFA algorithm and a CNN network, as it is shown in Fig.~\ref{PipeLine}.
\subsubsection{AEFA}
AEFA is a commonly used algorithm to compute automatically  a sufficient number of ellipses to model the given 2D human shape in the RGB image \cite{panagiotakis2016parameter}. As it is shown in Fig.~\ref{PipeLine}, AEFA provides a ground truth of body parts to train the CNN network.
\subsubsection{CNN Architecture}
A CNN uses a shared convolution kernel to process data and is commonly used for image recognition/classification.  
The CNN architecture used in this paper \cite{9410980} includes three convolutional pooling layers, each with a $3\times3$ convolution kernel, and with depths 32, 64, and 128, respectively. There is a pooling layer followed by each convolution layer, and the size of the pooling operation is $2\times2$. The pooling layer can compress the parameters of the network and reduce the over-fitting situation to a certain extent. The dropout layers are added to avoid overfitting. After convolution and pooling operations, data is flattened to a 1D vector and then is sent to the fully connected (FC) layers with one hidden layer, the neurons in the fully connected layers have connections with all the neurons in the previous layer, and are used to calculate the probability of the category. The hidden layer of the FC layers has a size of 128 neurons. The last part of the proposed CNN system is the output layer with 51 neurons, representing the 3D coordinates of key points. The ``ReLU'' function is used as the activation function for the layers. Finally, classification is performed using the “Softmax'' function. A learning rate of 0.001 and a mini-batch of 100 are used. The CNN architecture is selected to minimize the computational complexity and retain the classification accuracy.

The learning mechanism of the CNN is the same as that of the artificial neural network, which consists of the forward pass and backward pass. Specifically, we input the AEFA modeling result to the network and obtain the output of the CNN in the forward pass phase. Suppose there are $N_p$ predicted key points. We can leverage the difference between the predicted key point and the ground truth in the same frame to compute the mean square error (MSE). The $l_2$-norm distance between the $i$th estimated key point at frame $t$ and the corresponding ground truth is defined by: 
\begin{equation}
    \epsilon_t(i) = \frac{1}{2}\Vert \Tilde{C}_{t}^{i} - C_{t}^{i} \Vert_{2}^{2}, 
\end{equation}
where $\Tilde{C}_{t}^{i}$ represents the estimated $i$th key point coordinates at  frame $t$, while $C_{t}^{i}$ is the corresponding ground truth data. The loss function for the CNN at  frame $t$ is taken as: 
\begin{equation}
   \epsilon_t= \frac{1}{N_p}\sum_{i=1}^{N_p}\epsilon_t(i).
\end{equation} 
In the backward pass stage, the MSE is propagated along the network link path backward and the weights are updated along the negative gradient direction of the error function to minimize MSE.

\subsection{Human Vital Sign Estimation}

Due to the distance of a human to the radar, RBM may deteriorate the accuracy of the extraction of vital signs \cite{dai2021enhancement}. However, the MIMO radar at hand has multiple channels that can be used for beamforming towards the human chest. As it is shown in Fig.~\ref{PipeLine}, the AEFA-CNN algorithm can estimate the location of the chest, then after beamforming, vital signs are extracted. The process of extracting vital signs is as below: 
  \subsubsection{Range Bin Selection}
First, the range bin where the human is located is detected by applying an FFT in the fast time domain, where the size of FFT equals to the number of ADC samples in the radar. It is assumed that the target under test is the dominant scatterer in the investigation domain. As the radar has a full front view of the human in our case, a single range bin, that corresponds to maximum power, is selected.
 \subsubsection{Phase DC Compensation}
Due to the possible DC offset of the I/Q channels in the chirps at each frame, a DC compensation algorithm is applied on the chirp domain by non-linear least square as in \cite{alizadeh2019remote}.
  \subsubsection{Beamforming Towards Chest}
  The trained CNN algorithm estimates the torso and its center based on the point clouds from 50 frames. It is assumed that, by beamforming the recorded MIMO channels of the radar towards the chest, we can achieve a better vital sign extraction performance. Conventional analog beamforming is applied as:
  \begin{equation}
     d_s = \mathbf{y}(i_{chirp},:)\mathbf{v}_{chest}^*,
  \end{equation}
  where $\mathbf{y}(i_{chirp},:)\in \mathbb{C}^{1\times N_\text{MIMO}}$ denotes the vector of MIMO channels at the selected range bin and the beamforming is applied on all individual $i_{chirp}$ chirps. $\mathbf{v}_{chest}\in \mathbb{C}^{1\times N_\text{MIMO}}$ is the 2D steering vector toward the angle of chest, and $()^*$ denotes the complex conjugate operator.  $d_s$ is the resultant complex number at each chirp, and its phase contains information about vital signs.
\subsubsection{Phase Extraction}
As the displacements of the chest due to breathing are in the range of 0.01-12 mm which can cause the measured phase to exceed $\pm \pi$, such phase shifts are unwrapped. Since the harmonics of breathing are often dominant, a phase difference on the consecutive samples of phases, i.e., $\phi[k]-\phi[k-1]$ are applied, which suppresses the harmonics of breathing and allows the heart harmonic to be detected \cite{wang2020remote}.  
\subsubsection{Bandpass Filtering}
The BR and HR of healthy non-exercising adults are in  [0.1-0.5] Hz  and [0.8-1.7] Hz frequency bands, respectively \cite{hill2020monitoring,avram2019real}. Since the breathing and heart harmonics are the strongest components in each band, we apply a Butterworth filter of 5th-order at each band and apply an FFT to extract the dominant harmonics, which correspond to the vital signs. 
\begin{figure}
	\subfigure[]
	{\includegraphics[width=0.13\linewidth]{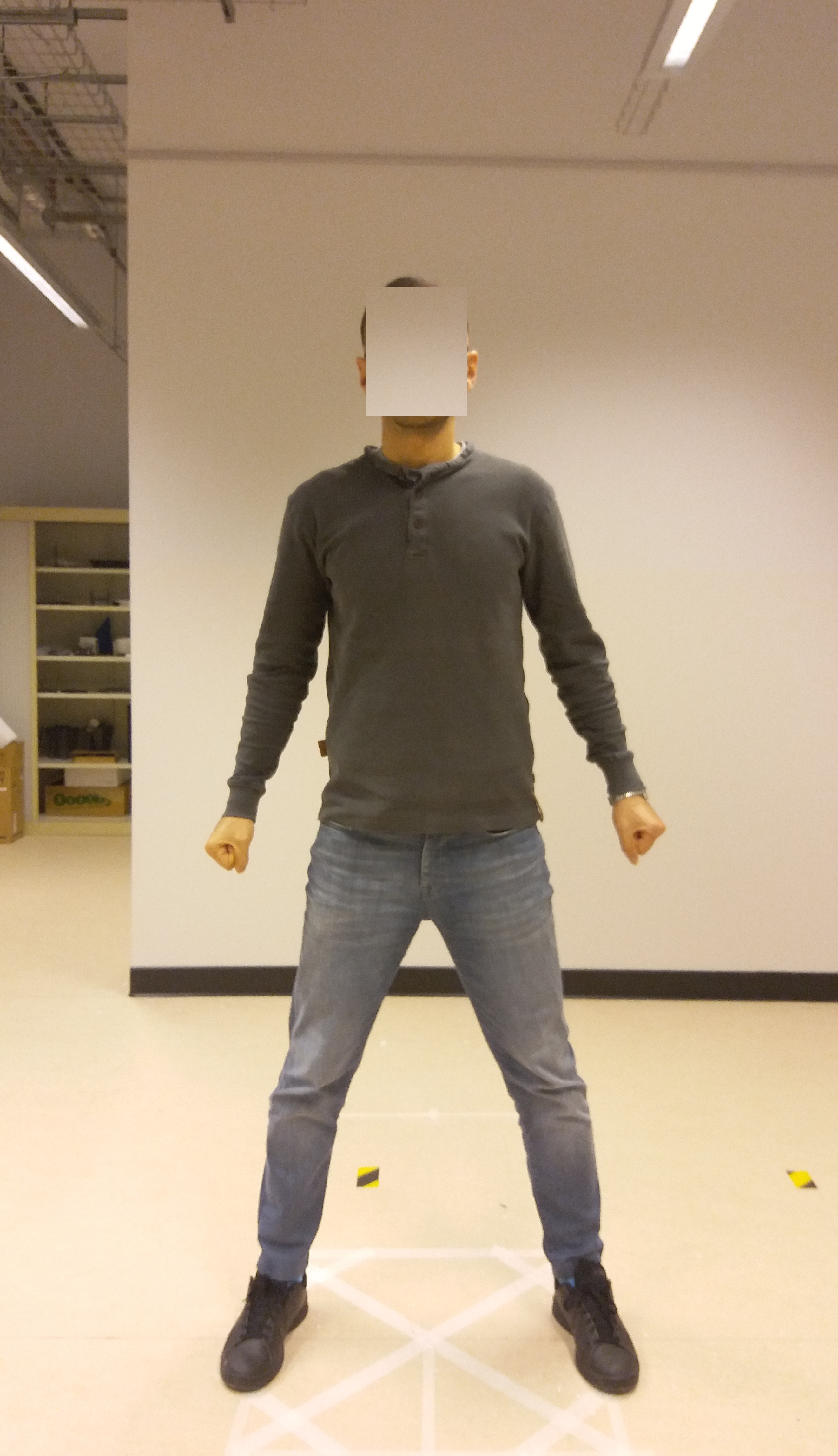}}
	\subfigure[]
   {\includegraphics[width=0.4\linewidth]{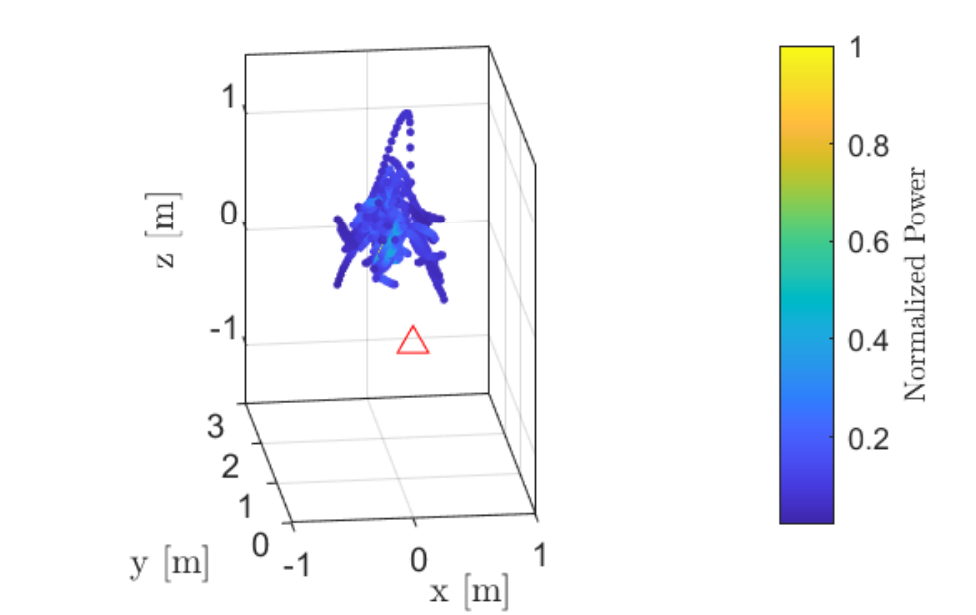}}
   \subfigure[]
   {\includegraphics[width=0.4\linewidth]{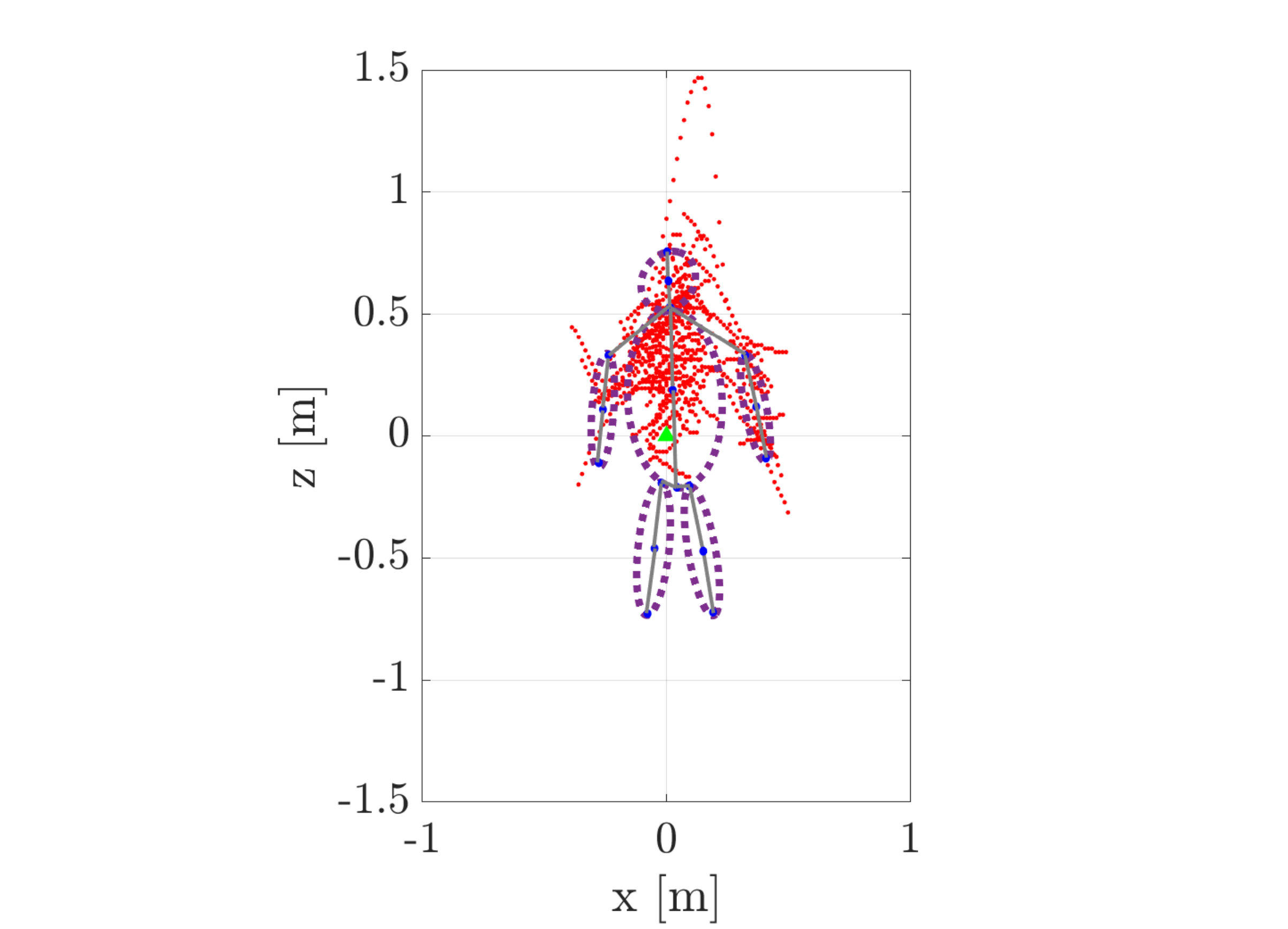}}\\
   \subfigure[]
	{\includegraphics[width=0.14\linewidth]{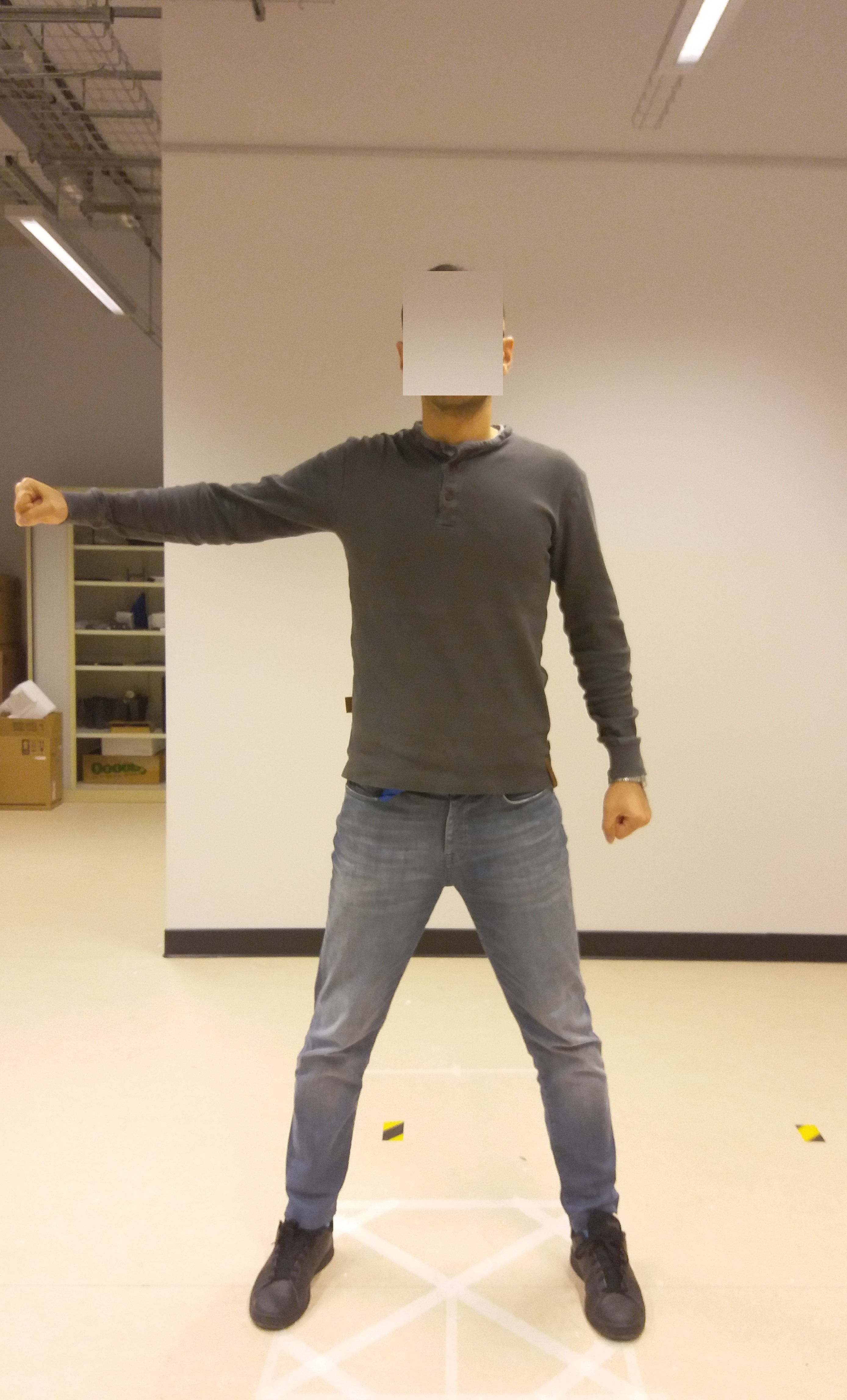}}
	\subfigure[]
   {\includegraphics[width=0.4\linewidth]{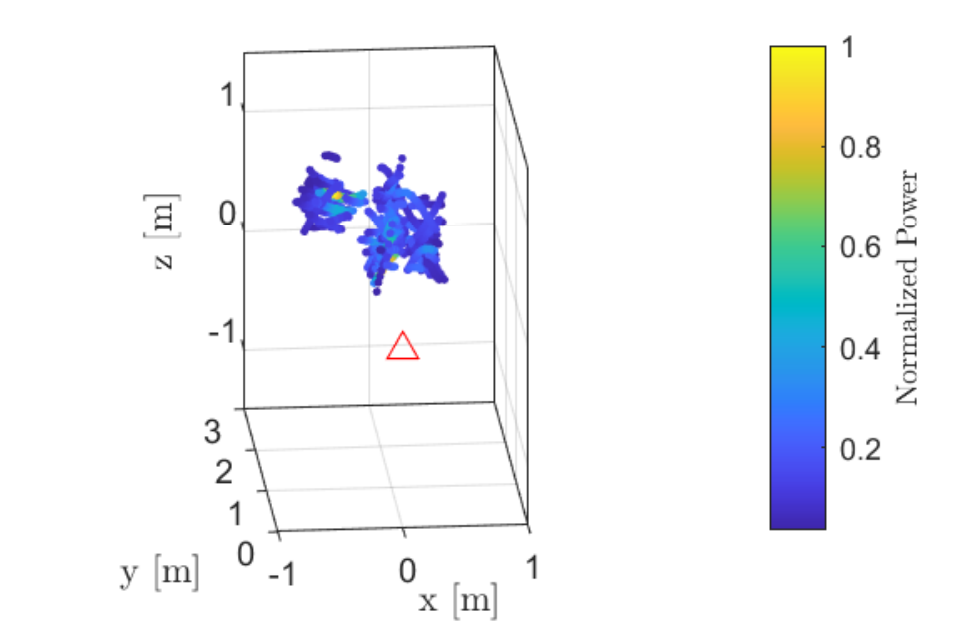}}
   \subfigure[]
   {\includegraphics[width=0.4\linewidth]{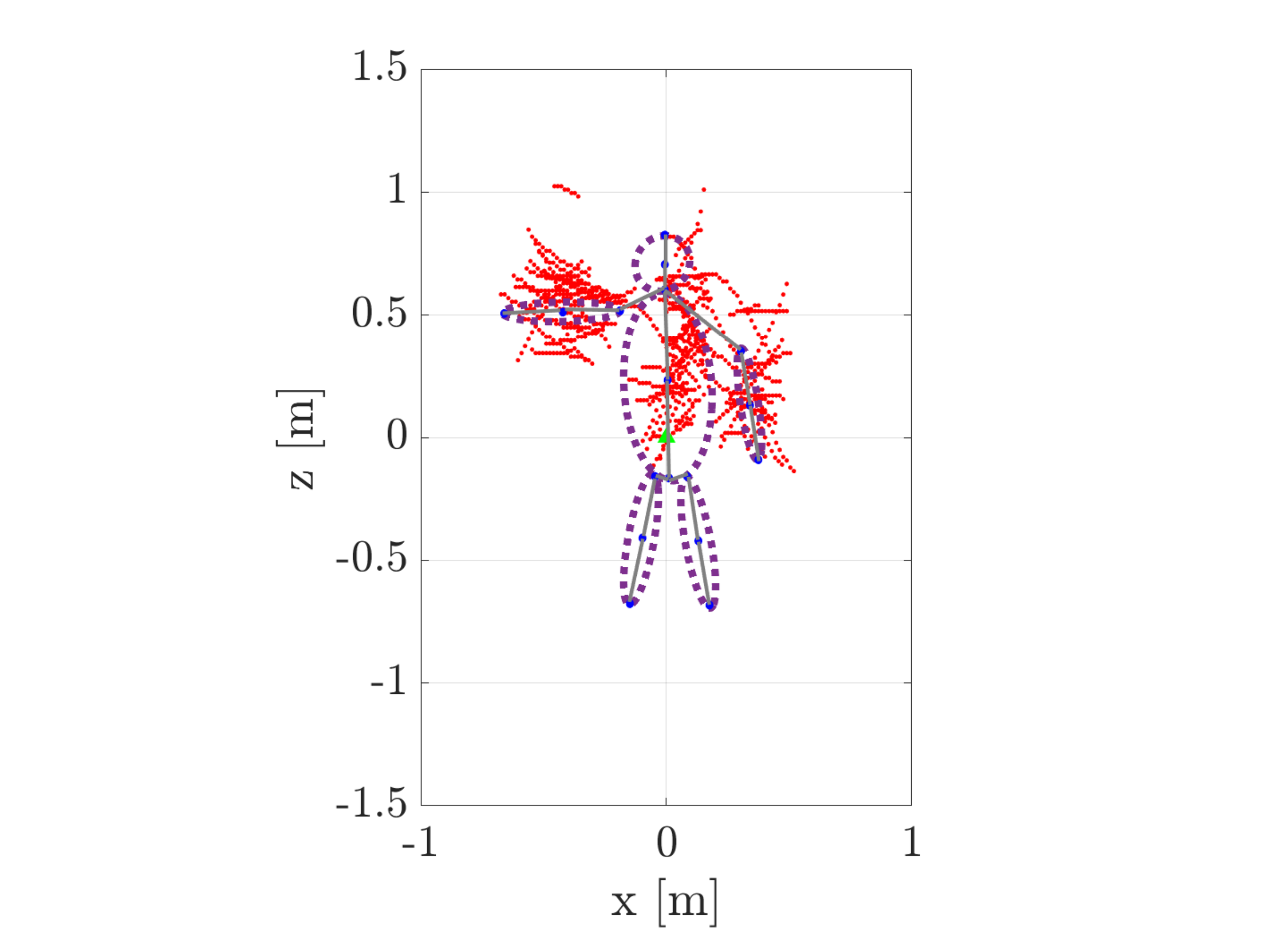}}\\
   \subfigure[]
	{\includegraphics[width=0.14\linewidth]{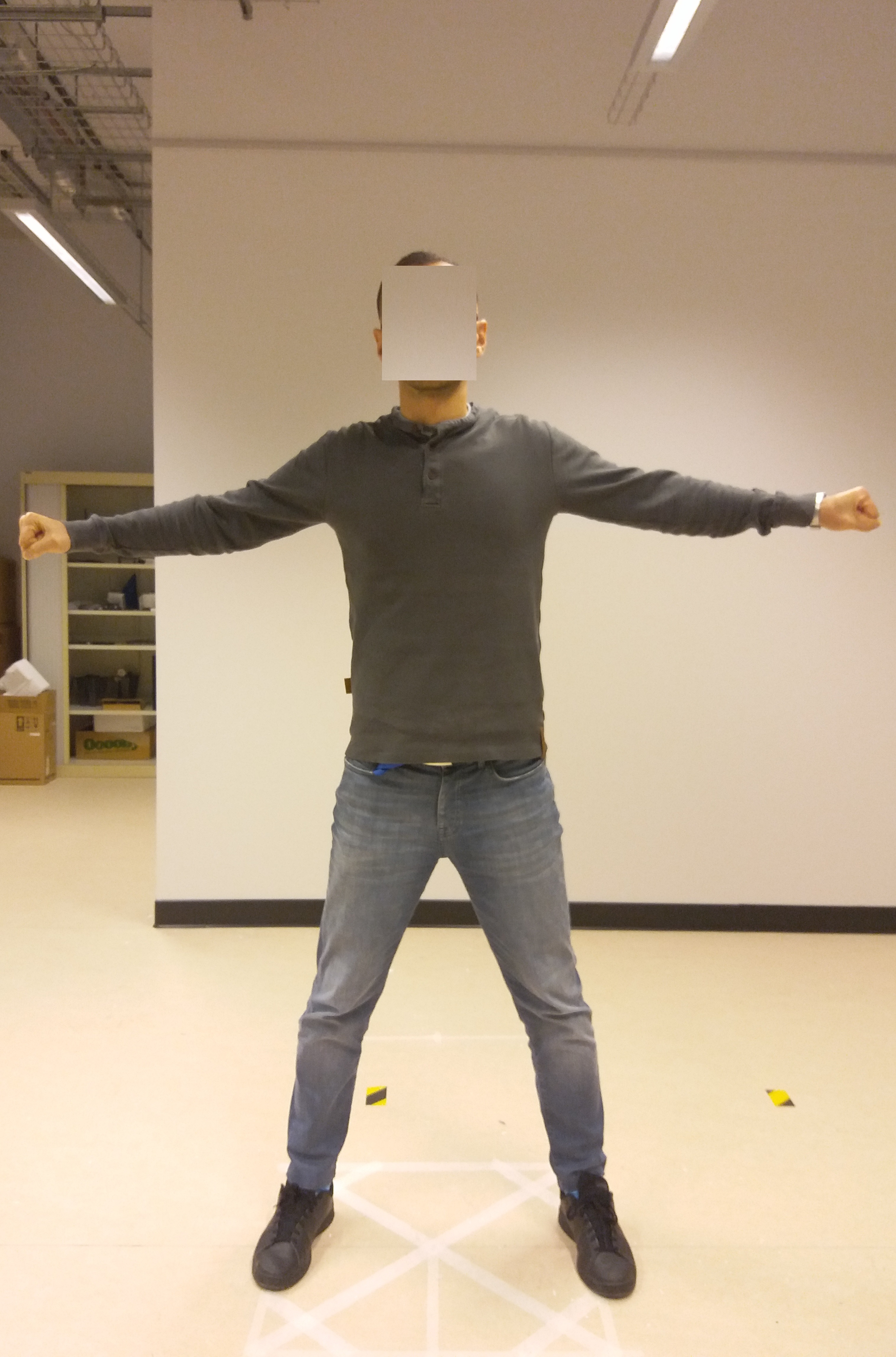}}
	\subfigure[]
   {\includegraphics[width=0.4\linewidth]{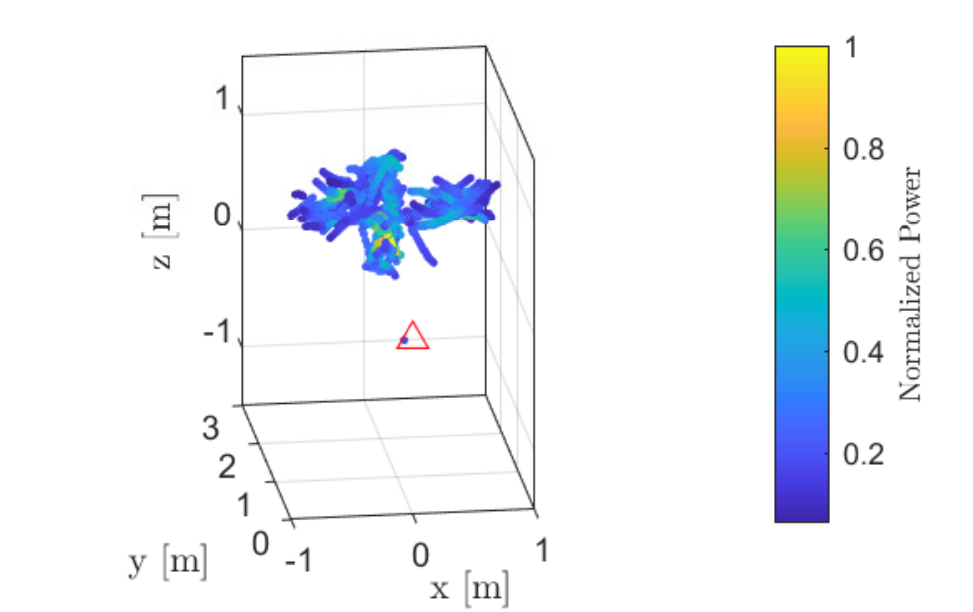}}
   \subfigure[]
   {\includegraphics[width=0.4\linewidth]{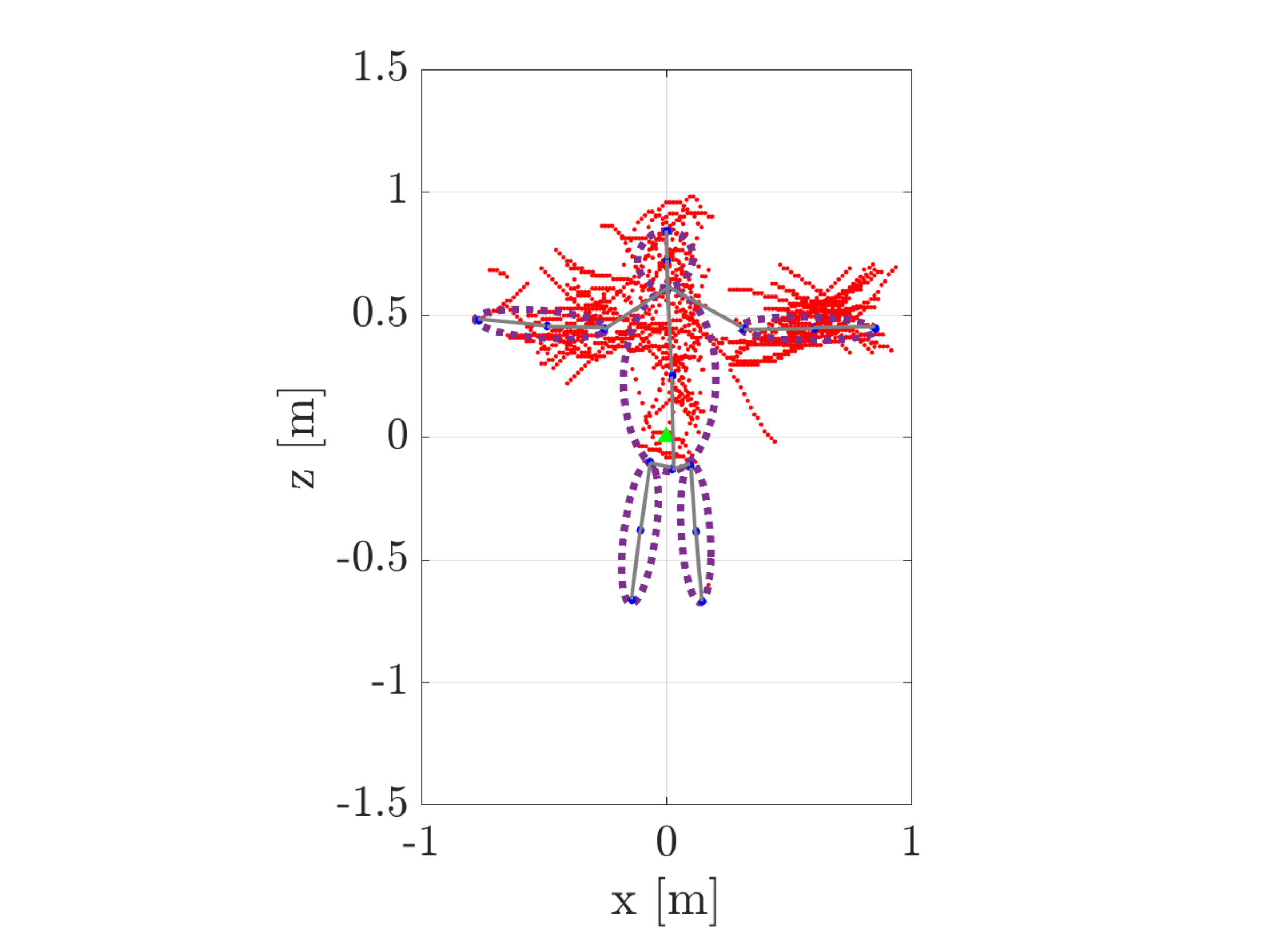}}
	\centering
	\caption{Human, point clouds and estimated shape by AEFA-CNN, (a-c) both-arms-down,  (a-c) one-arm-raised,  and (g-i) both-arms-raised. The red triangle shows the location of the radar.}
	\label{human}
\end{figure}
\begin{figure*}
    \subfigure[]	{\includegraphics[width=0.085\linewidth]{BothHandsDown.jpg}}
	\subfigure[]
	{\includegraphics[width=0.11\linewidth]{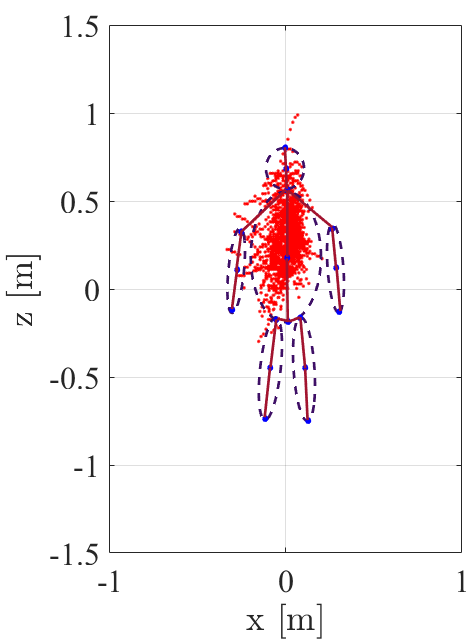}}
	\subfigure[]
	{\includegraphics[width=0.26\linewidth]{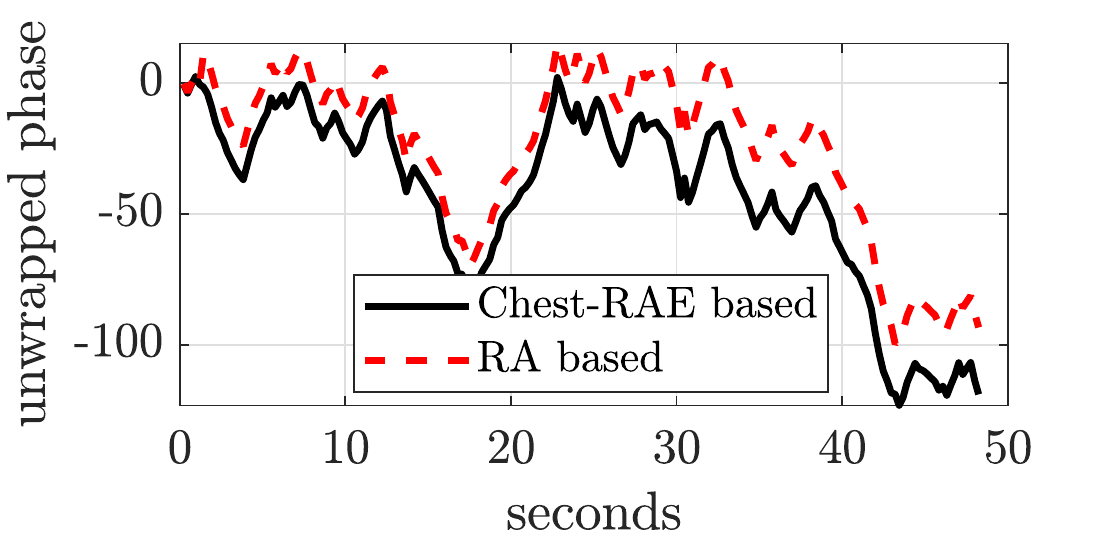}}
	\subfigure[]
   {\includegraphics[width=0.26\linewidth]{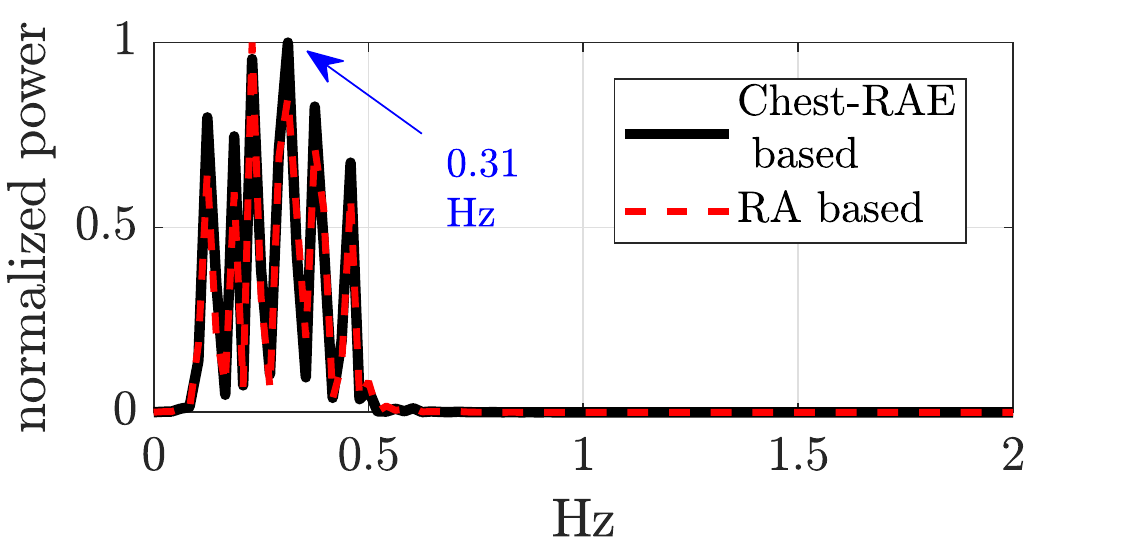}}
	   \subfigure[]
   {\includegraphics[width=0.26\linewidth]{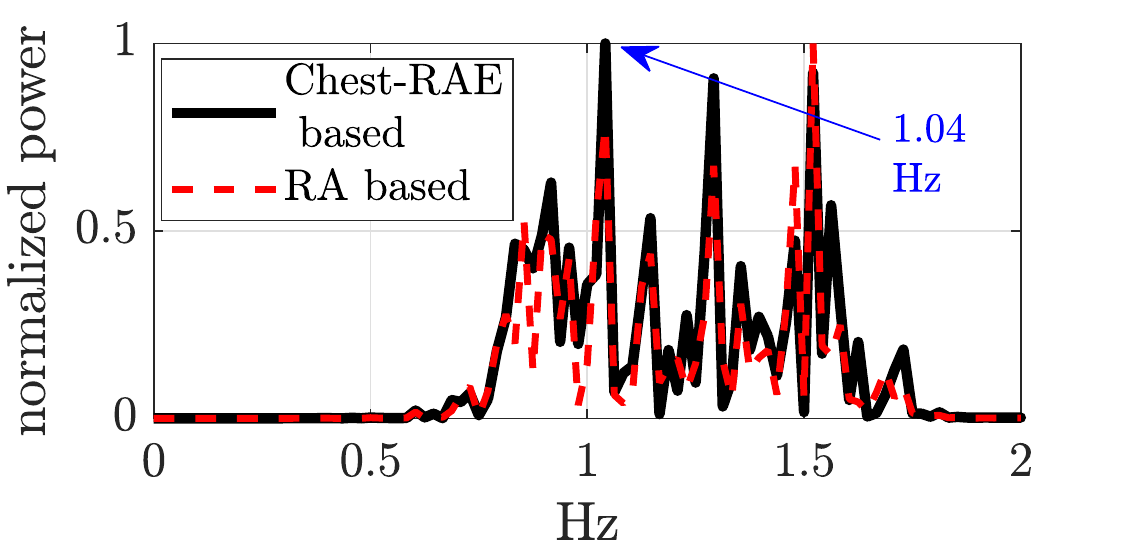}}
   \\
       \subfigure[]
	{\includegraphics[width=0.085\linewidth]{Human_OneHandRaised.jpg}}
	\subfigure[]
	{\includegraphics[width=0.11\linewidth]{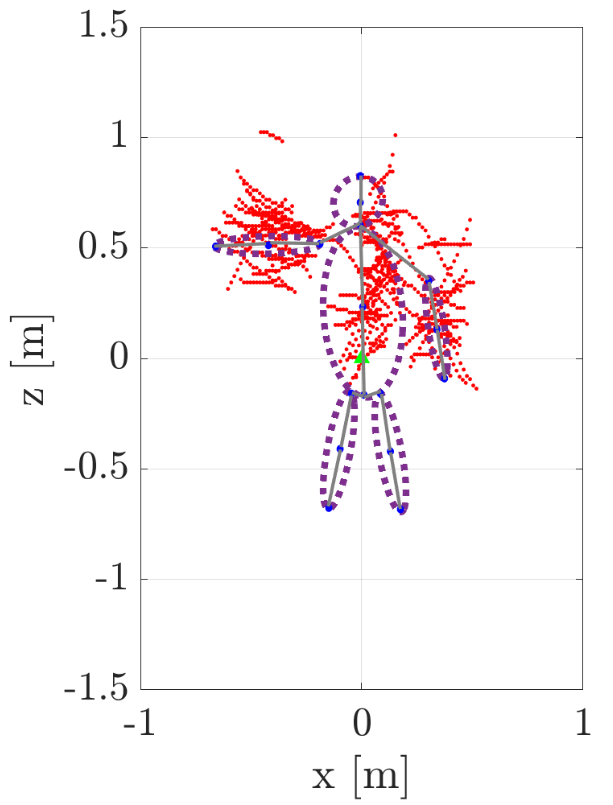}}
	\subfigure[]
	{\includegraphics[width=0.26\linewidth]{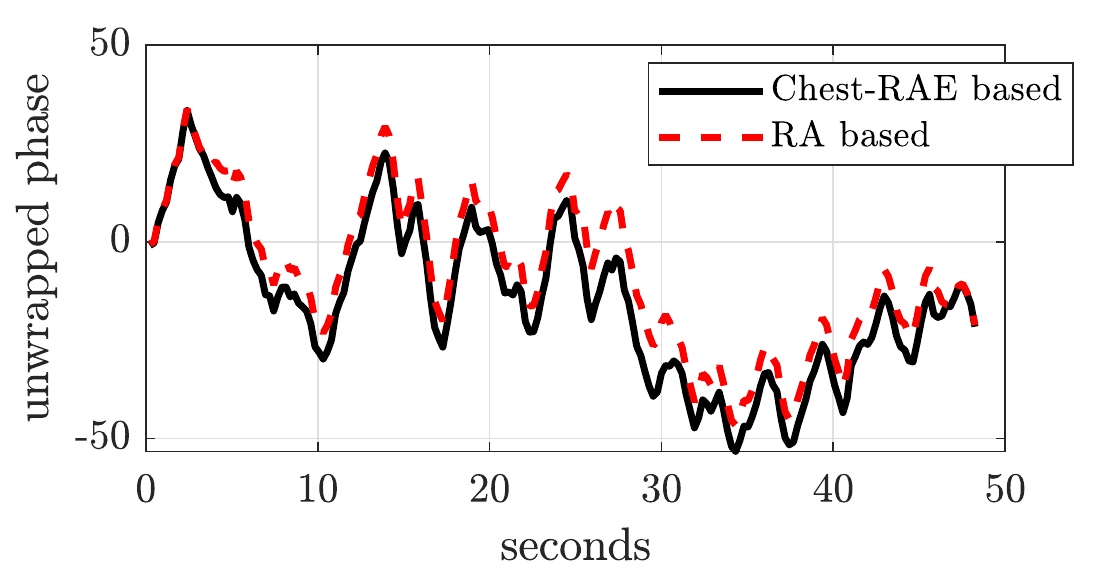}}
	\subfigure[]
   {\includegraphics[width=0.26\linewidth]{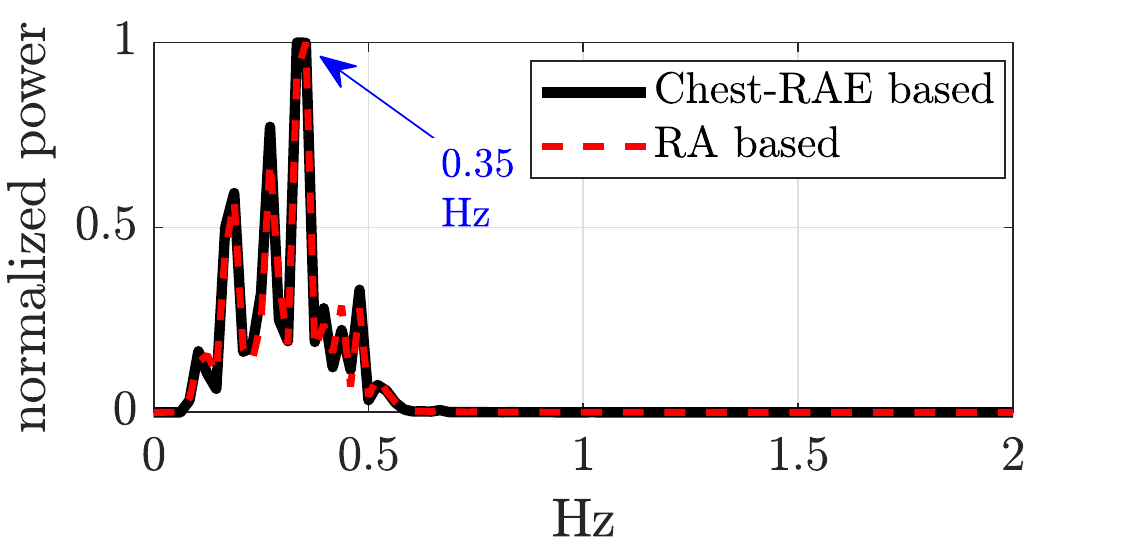}}
	   \subfigure[]
   {\includegraphics[width=0.26\linewidth]{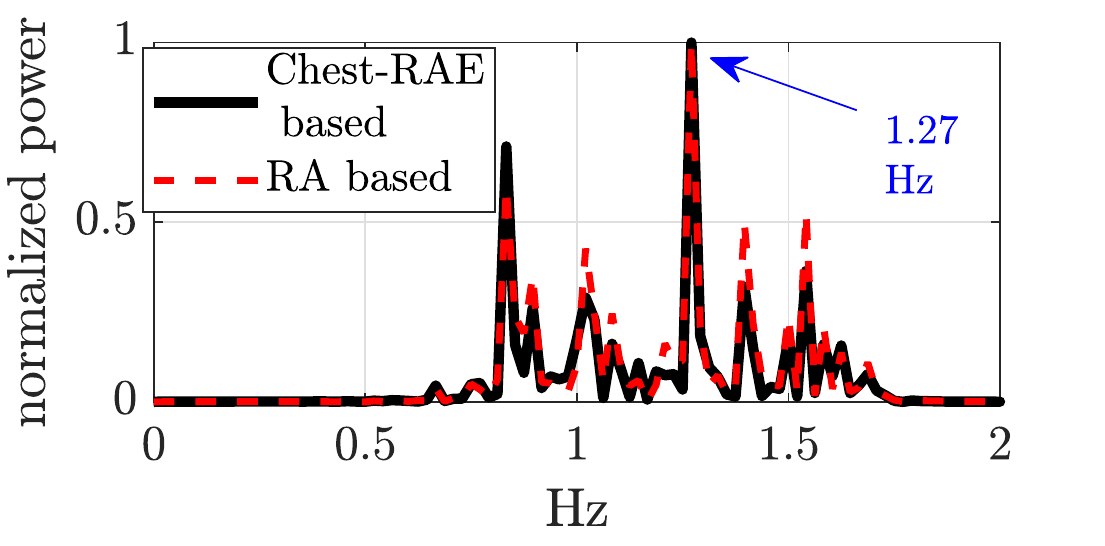}}
\\
	\subfigure[]	
 {\includegraphics[width=0.092\linewidth]{Human_bothArmsRaised.jpg}}
		\subfigure[]
	{\includegraphics[width=0.10\linewidth]{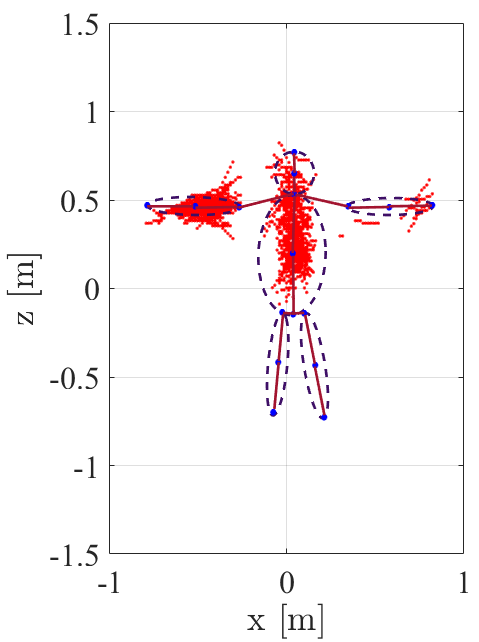}}
	\subfigure[]
	{\includegraphics[width=0.26\linewidth]{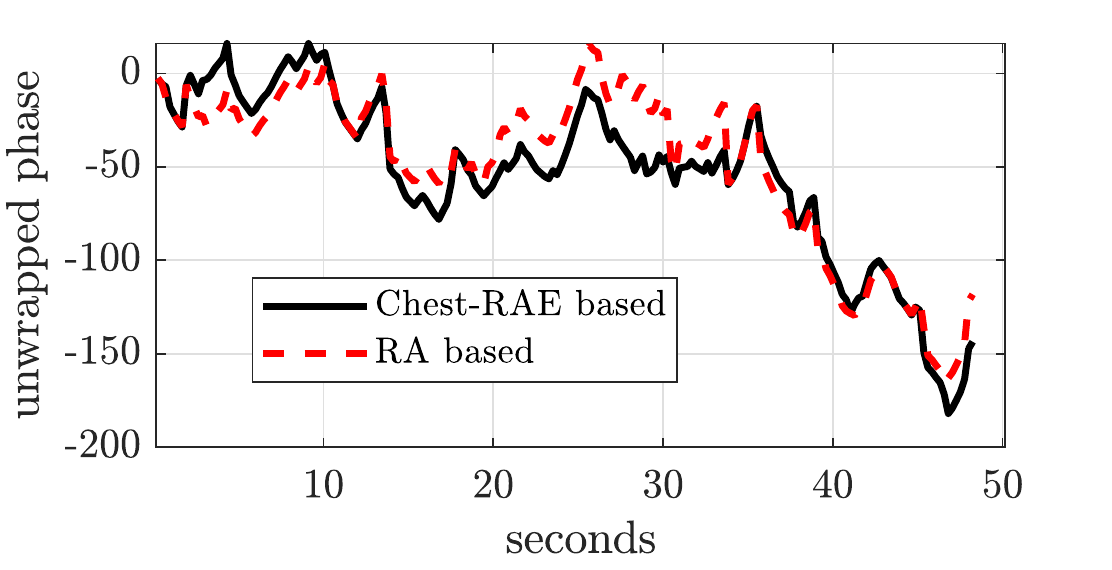}}
	\subfigure[]
   {\includegraphics[width=0.26\linewidth]{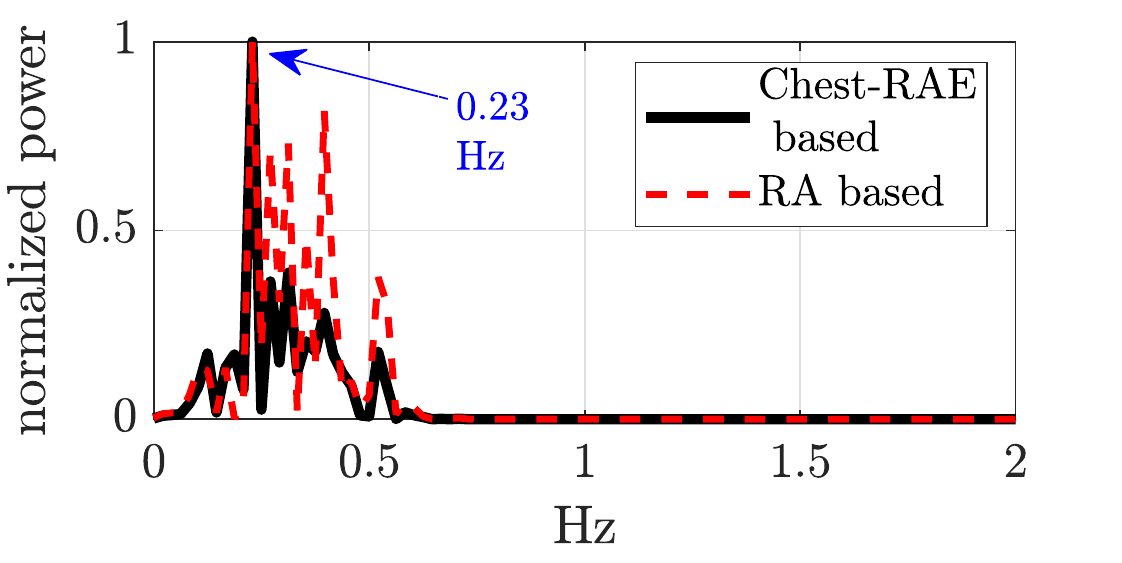}}
	\subfigure[]
   {\includegraphics[width=0.26\linewidth]{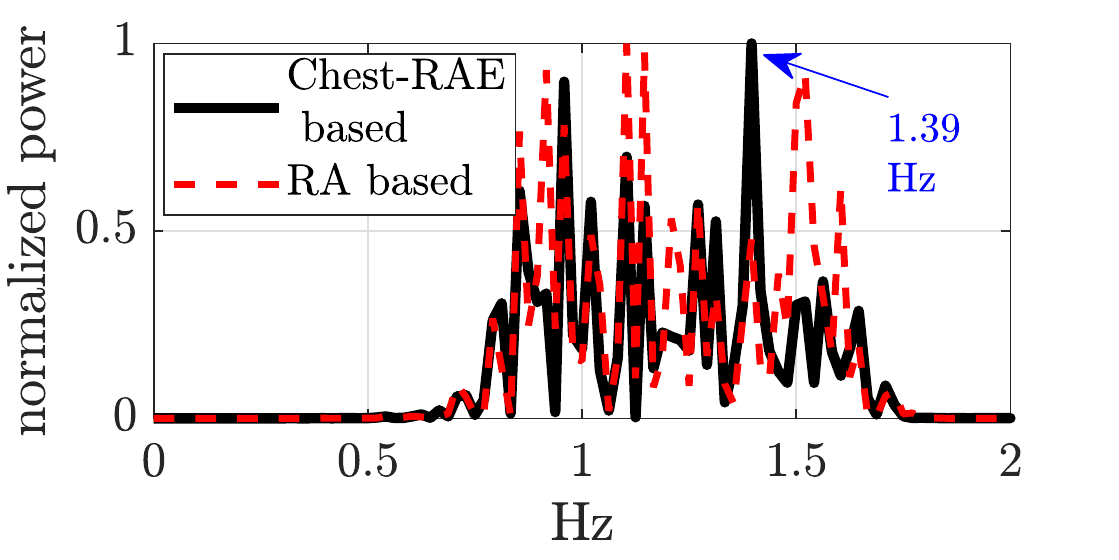}}
	\centering
	\caption{From left to right respectively, the human,  point clouds, extracted phase,  BR and HR estimation for  (a-e) BHD, (f-j) OHR and (k-o) BHR.}  
	\label{VitalSign}
\end{figure*}
\begin{table}
\centering
\begin{tabular}{|ll|ll|}
  \hline   
  Parameters & Value &  Parameters & Value \\
  \hline \hline
  Start Frequency & 77 GHz  & Sample Rate  & 2200kbps  \\
  \hline
   Chirp Rate & 60 MHz$/\mu$s & ADC Samples & 64  \\
  \hline
  Idle Time & 250$\mu$s & Bandwidth (BW) & 3.6 GHz \\
   \hline
   ADC start time & 10 $\mu$s &  Frame Duration &  240 ms\\
   \hline
   Ramp End Time  & 60 $\mu$s   & Chirps$/$frame  & 256 \\
  \hline
\end{tabular}
\caption{The parameters assigned for radar. }
\end{table}
\section{Implementation and Analysis}
\subsection{Measurement Scenario}
The radar utilized is Texas Instrument (TI) mm-wave FMCW chip (AWR1843) which includes 3 transmitting and 4 receiving antennas, and operates in time-division multiplexing MIMO manner at 76-81 GHz \cite{texas}. It is adjusted to $h=1.06$ m height from ground and, illuminating the full front view of a human in $r=2$ m distance of the radar as in the scenario in Fig.~\ref{scenario}. The details of the algorithms and parameters to produce point clouds are based on Table I and paper  \cite{abedi2021ai}. The range resolution can be adjusted to $4$ cm at a $3.6$ GHz bandwidth. However, the maximum resolutions in the azimuth and elevation of this radar are  $\Delta \phi=14^\circ$ and  $\Delta \theta=60^\circ$, respectively.  
Considering  the hardware specifics,  the resolutions are $\Delta x=r\Delta \phi=24$ cm and $\Delta z=r\Delta \theta=102$ cm, in $x$ and $z$-axes, respectively.  The purpose of this paper is to study only a standing human with variant arm postures, therefore these resolutions are sufficient.
A two-dimensional cell-averaging Constant False Alarm Rate (CFAR) algorithm is used to assign detections within the range-azimuth domain, and these are used to form a three-dimensional spatial point cloud. Through analyzing measurements obtained whilst illuminating a point target (small metallic sphere) the CFAR training  and guard band size were set to $[n_{g,x},n_{g,y}]=[8,8]$ and $[w_{g,x},w_{g,y}]=[8,8]$ respectively and the threshold was set to 10 dB after statistical analysis of the system noise floor.
After  applying 2D CFAR  on the range-azimuth map, the elevations of the detected points are calculated by the Capon algorithm. A frame duration of $T_f=240$ ms is assigned to generate enough point clouds at each frame to train the CNN.
The  postures, namely, both-arms-down (BAD),  one-arm-raised (OAR), and both-arms-raised (BAR) are chosen as the case studies, where  Fig.~\ref{human} shows point clouds of 50 frames combined and the estimated shape by AEFA-CNN, which is trained by the point clouds of 150 frames beforehand.

 Considering a maximum frequency of interest $2$ Hz for the harmonics in vital signs, the sampling frequency should be $f_s> 4$ Hz, therefore  the sampling time of phase signals which equals the frame duration is kept as $T_f=240$ ms. 
A human in three postures, namely BAD, OAR, and BAR, is studied. The radar recorded 350 frames, of which the first 150 frames are utilized for the training of the AEFA-CNN,  and then vital signs are extracted from the remaining frames. To show the effectiveness of estimating the location of the chest in the azimuth and  elevation domains on improving vital sign estimation, we demonstrate the vital sign extraction by two approaches, namely range-azimuth (RA)-based where the range and azimuth angle of human are considered, and the range-azimuth-elevation (RAE)-based which the chest is targeted for phase extraction. The ground truths in HR estimations are based on the measurement by Apple Watch for one minute, which are 1.1, 1.25, and 1.41 Hz for BAD, OAR, and BAR, respectively.
Fig.~\ref{VitalSign} shows the human with three arm postures and the estimated shapes by the CNN.
  We define peak-to-average power ratio (PAPR) as the peak power divided by the average power of the harmonics in vital signs, and compare the results from RA- and RAE-based beamforming. The effect of beamforming toward the chest is demonstrated in BR and HR estimation where chest-RAE-based vital sign extraction provides fewer artifacts, i.e., Fig.~\ref{VitalSign}(e), (j) and (o) when estimating HR,  Fig.~\ref{VitalSign}(n) when estimating BR.  
  The improvements in PAPR are better observed in Fig.~\ref{VitalSign} (j) 0.02 dB and  Fig.~\ref{VitalSign}(n)  with 0.07 dB, while both methods could correctly estimate the rates of vital signs. In Figs.~\ref{VitalSign} (e) and  (o), RA based method even fails to determine the true HR rate.
The increase of HR from  1.04 Hz for BHD to 1.41 Hz for BHR is also observed due to the activity of heart to keep the arms raised.  

The pipeline proposed here is a single person case study, which includes a full-front view of the human. A radar with a higher resolution in elevation can intensify the effect of beamforming toward the chest, which is observed through PAPR of the harmonics  in the vital signal.

\section{Conclusion}
The benefit of human posture estimation for vital sign estimation is demonstrated in this paper.
The localization of the chest followed by beamforming can improve the estimation of heart  and breathing rates, since this spatial filtering reduces the interference caused by non-informative body motions. In the selected range bin of a human in front of a radar, we compare the vital sign estimation by 1D beamforming to the azimuth angle of the human  with the 2D beamforming towards the azimuth and elevation of the chest. The results show that when the local peaks occur at the same frequencies for both 1D and 2D beamforming, 2D beamforming indicates higher peak-to-average power ratio in the harmonics of vital signal, which is beneficial for vital sign estimation.
An in-depth study of rotated human and various distances of radar and human are intended for our next work.

\bibliographystyle{IEEEtran}

\end{document}